\documentclass[prl,aps,twocolumn,superscriptaddress,showpacs]{revtex4}
\usepackage{epsfig,amsmath}
\begin{document}

\title{Conditional preparation of maximal polarization entanglement}

\author{Cezary \'Sliwa}
\affiliation{Clarendon Laboratory,
University of Oxford, Parks Road, Oxford OX1 3PU, United Kingdom}
\affiliation{Centre for Theoretical Physics,
Al.\ Lotnik\'ow 32/46, 02-668 Warsaw, Poland}

\author{Konrad Banaszek}
\affiliation{Clarendon Laboratory,
University of Oxford, Parks Road, Oxford OX1 3PU, United Kingdom}

\date{\today}

\begin{abstract}
  A simple experimental setup consisting of a spontaneous parametric
  down-conversion source and passive linear optics is proposed for
  conditional preparation of a maximally entangled polarization state
  of two photons. Successful preparation
  is unambiguously heralded by coincident
  detection of four auxiliary photons. The proposed scheme utilizes the
down-conversion term corresponding to the generation of
three pairs of photons.
We analyze imperfect detection of the auxiliary photons
and demonstrate that its deleterious effect on the fidelity of the prepared
state can be suppressed at the cost of decreasing the efficiency of the scheme.
\end{abstract}

\pacs{03.65.Ud, 42.50.Dv, 03.67.-a}

\maketitle


Pairs of photons in a maximally entangled polarization state
constitute a basic constructional
primitive in many protocols for quantum information processing,
including teleportation, dense coding, and cryptography
\cite{ThePhysicsofQI}. They have also
been serving as a source of nonlocal correlations that violate Bell's
inequalities, thus contradicting assumptions underlying local realistic
theories. Despite enormous progress in generating entangled states of
photons \cite{KwiaMattPRL95},
starting from the initial experiments with atomic cascades \cite{AspeGranPRL82},
deterministic polarization entanglement in the photonic domain remains
an elusive entity \cite{KokBrauPRA00,KokBrauPRA00-2}.
Indeed, the majority of current
experiments is based on the production of photon pairs in the process
of spontaneous parametric down-conversion \cite{KwiaMattPRL95},
which is inherently random.
Consequently, it is possible to determine whether a pair has been
generated only by postselection, when looking {\em a posteriori} at
the number of detected photons. This property is not essential in some
applications such as tests of Bell's inequalities, but it becomes
critical especially in experiments involving multiple photon pairs
\cite{KokBrauPRA00,ProblemswithMultiplePairs}.
The random character of down-conversion sources may not be
shared in the future by the solid-state sources of single photons or
photon pairs that are presently being developed \cite{SolidStateSources},
though they will probably require operation at liquid-helium
temperatures.

From a practical point of view, spontaneous parametric
down-conversion in nonlinear crystals
is a stable and robust process
that requires modest experimental means to set up. An interesting
problem is therefore whether the randomness of the parametric sources
could be overcome by means of conditional detection. In such a scheme,
detecting a number of auxiliary photons by trigger detectors would
provide {\em a priori} information that an entangled photon pair has
been generated, without destructive photodetection. Such a pair could
be used in the event-ready manner, or possibly stored in a cavity
\cite{CavityStorage}
or an atomic system \cite{AtomicStorage}
for later use at any instant of time. The most
natural approach to realize this idea would be to perform the
procedure of entanglement swapping on two entangled pairs generated
independently, one by each of the two crystals.
The Bell measurement would then play a
twofold role of collapsing the state of the remaining photons onto
an entangled state as well as assuring their presence
\cite{ZukoZeilPRL93}. However, when
the pairs are generated in parametric down-conversion, it is necessary
to take into account other processes whose probability of occurrence is
of the same order of magnitude, such as generation of a double pair in
one crystal and none in the second crystal \cite{KokBrauPRA00}.
This turns out to be a fundamental obstacle
in the conditional preparation of maximal entanglement from four
down-converted photons: it has been shown \cite{KokBrauPRA00-2} that a
maximally entangled state cannot be generated with a nonzero
probability in any setup comprising down-converters and linear optics,
based on detection of two auxiliary photons. This rules out the possibility
of using the second-order term of the spontaneous down-conversion output,
containing overall four photons, to produce maximally entangled pairs by means
of conditional detection.

In this paper we show that the
six-photon component of the spontaneous down-conversion output suffices to
generate the maximally entangled polarization state of two photons in an
event-ready manner. Specifically,
we propose here an experimental setup based on a single nonlinear
crystal producing two beams containing photons with pairwise correlated
polarizations. We demonstrate that fourfold coincidence detection performed on
fractions of the output beams picked off with nonpolarizing beam splitters
leaves the remaining modes in a maximally entangled
two-photon state. The use of the third-order term of spontaneous
down-conversion places certainly more stringent requirements on
the brightness of the necessary sources, but as we discuss later
the present progress in down-conversion sources
is likely to make this idea feasible in the near future. The proposed setup
presents a substantial advancement over the scheme implied directly by the
general methodology of quantum computing with linear optics \cite{LOQC},
whose implementation with down-conversion sources would require in total
four photon pairs \cite{KokPhD}.

\begin{figure}[btp]
  \centering
  \includegraphics[width=0.8\columnwidth]{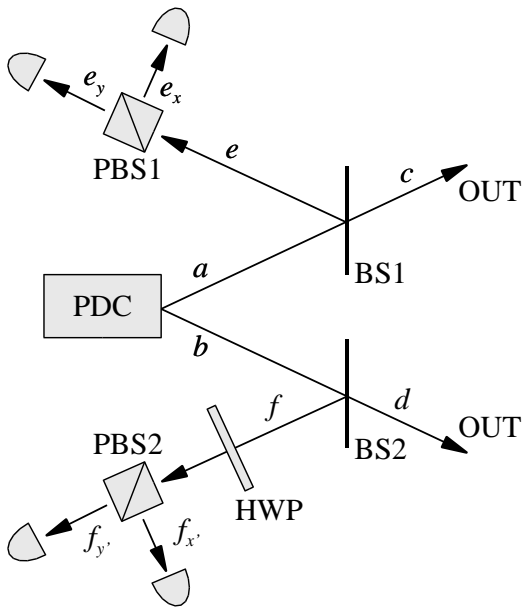}
  \caption{Schematic representation of the setup for generating
 heralded entanglement. BS1, BS2, nonpolarizing beam splitters;
HWP, half-wave plate; PBS1, PBS2, polarizing beam splitters. Lowercase
letters label the beams.}
  \label{fig:scheme}
\end{figure}

The proposed setup, shown in Fig.~\ref{fig:scheme},
consists of one pumped nonlinear
crystal generating pairs of down-converted photons entangled in their
polarizations, and a passive
optical circuit directing the down-converted photons into the output
and detected modes. We label the modes with the corresponding annihilation
operators. We assume that the conversion rate is low, which will
allow us to describe the down-conversion process using the perturbative expansion
in the number of the produced photons.
Let $(\hat{a}_x, \hat{a}_y; \hat{b}_x, \hat{b}_y)$
be the polarization
modes of the down-converted photons. The beam $(\hat{a}_x, \hat{a}_y)$ is directed
to a nonpolarizing
beam splitter BS1 with the amplitude transmission coefficient $\cos\theta_a$, where the
transmitted modes $(\hat{c}_x, \hat{c}_y)$ constitute the setup output,
while the reflected modes $(\hat{e}_x, \hat{e}_y)$ are detected locally. The other beam
$(\hat{b}_x, \hat{b}_y)$ is directed to another beam splitter BS2 with
the amplitude transmission coefficient
$\cos\theta_b$, where the transmitted modes $(\hat{d}_x, \hat{d}_y)$
are the remaining two output modes of the setup, while the reflected modes
$(\hat{f}_x, \hat{f}_y)$ are analyzed in the $\pi/4$ rotated polarization modes
$(\hat{f}_{x'}, \hat{f}_{y'})$, defined by
the relations $\hat{f}_{x}=(\hat{f}_{x'}+\hat{f}_{y'})/\sqrt{2}$ and
$\hat{f}_{y}=(\hat{f}_{x'}-\hat{f}_{y'})/\sqrt{2}$.

We assume that the photodetectors monitoring the auxiliary modes
can perform the ideal projection onto the one-photon Fock state
$|1\rangle\langle 1|$. We will be interested in events
when each of the four auxiliary detectors registers exactly one photon.
We shall demonstrate that the lowest order for which such an event can occur
is when three pairs of down-converted photons are generated, and
that in this order a fourfold coincident
detection of single photons in the auxiliary
modes $(\hat{e}_x, \hat{e}_y; \hat{f}_{x'}, \hat{f}_{y'})$ unambiguously
heralds that the quadruplet of the output modes
$(\hat{c}_x, \hat{c}_y; \hat{d}_x, \hat{d}_y)$ contains a pair of photons
in a maximally entangled polarization state. This procedure prepares
a maximally
entangled state without the usual vacuum contribution, being
consequently a source of event-ready entanglement in the photonic domain
\cite{ZukoZeilPRL93}.

Let us now discuss the operation of the scheme in quantitative terms. The Hamiltonian
governing the down-conversion process in the weak conversion regime describes two independent processes, corresponding to generation
or deletion of a photon pair in modes $(\hat{a}_x;\hat{b}_y)$ or $(\hat{a}_y;\hat{b}_x)$,
respectively. These two processes are added coherently with opposite probability amplitudes. Assuming that the effective dimensionless interaction time is $r$,
we can decompose the output state
into terms that contain a fixed number of down-converted photons
\cite{KokBrauPRA00,LamaHoweNAT01}:
\begin{eqnarray}
|\Psi\rangle & = & \exp[r ( \hat{a}_x^{\dagger} \hat b_y^{\dagger}
         -\hat{a}_x \hat{b}_y )
- r ( \hat{a}_y^{\dagger} \hat{b}_x^{\dagger} - \hat{a}_y \hat{b}_x )]
|\text{vac}\rangle
\nonumber \\
\label{Eq:Psi}
& = & \sum_{n=0}^{\infty} \lambda_n |\Psi_n \rangle,
\end{eqnarray}
where
\begin{equation}
\label{Eq:lambdan}
\lambda_n = \sqrt{n+1} \frac{\tanh^n r}{\cosh^2 r}
\end{equation}
is the probability amplitude of generating $n$ photon pairs, and
the normalized $n$-pair component of the wave function takes the following form:
\begin{eqnarray}
|\Psi_n \rangle & = & \frac{1}{\sqrt{n+1}} \sum_{m=0}^n (-1)^m
      | n-m, m; m, n-m \rangle
\nonumber \\
\label{Eq:Psin}
 & = & \frac{1}{n!\sqrt{n+1}} ( \hat a_x^{\dagger} \hat b_y^{\dagger}
        - \hat a_y^{\dagger} \hat b_x^{\dagger} )^n
    | \text{vac} \rangle.
\end{eqnarray}
The occupation numbers in the first expression
for $|\Psi_n \rangle$ correspond to the ordering of the modes as
$(\hat{a}_x, \hat{a}_y ; \hat{b}_x , \hat{b}_y )$.

Let us first show that any of the terms with $n<3$ cannot give a fourfold
coincidence on the auxiliary detectors. Obviously, the only term that could
possibly give rise to such an event corresponds to $n=2$ and is
explicitly given by
\begin{equation}
\label{Eq:Psi2}
\left| \Psi_2 \right> = \frac{1}{\sqrt{3}}
 ( \left| 2, 0; 0, 2 \right> - \left| 1, 1; 1, 1 \right>
 + \left| 0, 2; 2, 0 \right> ) .
\end{equation}
The fourfold coincidence event
implies that all the four photons have been reflected by the beam splitters
BS1 and BS2. However, detection of two photons in the modes $(\hat{e}_x, \hat{e}_y)$
means that we are observing the middle term of the sum in Eq.~(\ref{Eq:Psi2}),
and that the state of the remaining two photons is collapsed to 
$|1_{\hat{f}_{x}}, 1_{\hat{f}_{y}}\rangle$. This state is transformed by
the half-wave plate to the form 
$  ( |2_{\hat{f}_{x'}}, 0_{\hat{f}_{y'}} \rangle
    - |0_{\hat{f}_{x'}}, 2_{\hat{f}_{y'}} \rangle )/\sqrt{2}$,
which of course cannot give a coincidence on the detectors monitoring the
modes $\hat{f}_{x'}$ and $\hat{f}_{y'}$. This is essentially the destructive
two-photon interference effect observed first by
Hong, Ou, and Mandel \cite{HongOuPRL87}.

Having shown that $n=3$ is the lowest order that can contribute to the fourfold
coincidence event in our scheme, let us now find the conditional state of the output
modes provided that each of the auxiliary detectors has seen exactly one photon.
The easiest way to approach this task is to use the second expression for the
state $|\Psi_3\rangle$ given in Eq.~(\ref{Eq:Psin}) in terms of the creation operators:
\begin{equation}
\label{Eq:Psi3}
|\Psi_3\rangle = \frac{1}{12} 
( \hat a_x^{\dagger} \hat b_y^{\dagger}
        - \hat a_y^{\dagger} \hat b_x^{\dagger} )^3
    | \text{vac} \rangle.
\end{equation}
The linear optics placed after the nonlinear crystal is described by the following
transformation of the annihilation operators of the down-conversion
modes:
\begin{subequations}
\begin{eqnarray}
\label{Eq:axassum}
\hat{a}_x & = & \hat{c}_x \cos\theta_a + \hat{e}_x \sin\theta_a,
\\
\hat{a}_y & = &  \hat{c}_y \cos\theta_a + \hat{e}_y \sin\theta_a,
\\
\hat{b}_x & = & \hat{d}_x \cos\theta_b + 
(\hat{f}_{x'} + \hat{f}_{y'})\sin\theta_b/\sqrt{2},
\\
\hat{b}_y & = & \hat{d}_y \cos\theta_b +
(\hat{f}_{x'} - \hat{f}_{y'})\sin\theta_b/\sqrt{2}.
\end{eqnarray}
\end{subequations}
After
inserting the above representation into Eq.~(\ref{Eq:Psi3}), we 
expand the resulting polynomial and isolate terms that contribute
to the coincidence event of interest.
These terms must contain the combination of
the creation operators of the auxiliary modes in the form $\hat e_x^{\dagger} \hat e_y^{\dagger}\hat f_{x'}^{\dagger} \hat f_{y'}^{\dagger}$. 
The above rather lengthy procedure can be aided with a computer algebra system.
Explicitly, the relevant
terms are
\begin{eqnarray}
  | \Psi_3 \rangle  & = &
    \frac{1}{\sqrt{2}} 
      \sin^2\theta_a \cos\theta_a \sin^2\theta_b \cos\theta_b
  \nonumber \\
& &
\times
    \frac{1}{\sqrt{2}}( \hat c_x^{\dagger} \hat d_x^{\dagger}
       + \hat c_y^{\dagger} \hat d_y^{\dagger})
      \hat e_x^{\dagger} \hat e_y^{\dagger}
      \hat f_{x'}^{\dagger} \hat f_{y'}^{\dagger}
    | \text{vac} \rangle + \ldots,
\end{eqnarray}
where ``$\ldots$'' denote all the other components. It is thus seen that the coincident
detection of four photons in each of the modes $\hat{e}_x, \ldots , \hat{f}_{y'}$
nondestructively
collapses the state of the output modes to the vector 
\begin{equation}
\label{Eq:Phi}
|\Phi\rangle = \frac{1}{\sqrt{2}}( \hat c_x^{\dagger} \hat d_x^{\dagger} + \hat c_y^{\dagger} \hat d_y^{\dagger}) | \text{vac} \rangle,
\end{equation}
which describes a maximally entangled state of two photons. This state can be
of course transformed into any other Bell state with the help of phase shifters
and polarization rotators.

The efficiency of successful state preparation in our scheme can be defined
as the probability of the fourfold coincidence event assuming that exactly
three photon pairs have been generated in the down-conversion process.
It is explicitly given by
\begin{equation}
  P = \frac{1}{2}
    (\sin^2\theta_a \cos\theta_a \sin^2\theta_b \cos\theta_b)^2,
\end{equation}
and it attains its maximum, equal to $(2/9)^3 \approx 0.011$,
when the power transmission coefficients of the beam splitters are
$\cos^2\theta_a = \cos^2\theta_b = 1/3$. In order to obtain the overall
preparation efficiency, this value of $P$
needs to be multiplied by the probability $(\lambda_3)^2$, defined
by Eq.~(\ref{Eq:lambdan}), of generating
the six-photon state. As it would be beneficial to run the down-conversion
process with the interaction parameter $r$ as large as possible while staying
within the perturbative regime, we have also estimated the contribution from
the next leading order term given by the state $|\Psi_4\rangle$. The probability
of producing a fourfold single-photon coincidence by this component is
$\frac{13}{5}(\sin\theta_a\cos\theta_a\sin\theta_b\cos\theta_b)^4$, which for
the choice of the transmission coefficients optimizing $P$ gives approximately
$6.3 \times 10^{-3}$.

The operation of the proposed scheme can be understood more intuitively
using the Fock state representation of the state $|\Psi_3\rangle$:
\[
\left| \Psi_3 \right> = \frac{1}{2}
 ( | 3, 0; 0, 3 \rangle - | 2, 1; 1, 2 \rangle
 + | 1, 2; 2, 1 \rangle - | 0, 3 ; 3 , 0 \rangle) .
\]
Observation of a twofold coincidence in the modes $(\hat{e}_x, \hat{e}_y)$ means
that one photon has been extracted from each of the modes $\hat{a}_x$ and
$\hat{a}_y$. This leaves us with the state of the remaining photons
proportional to $-|1,0;1,2\rangle + |0,1;2,1\rangle$. Further, a twofold
coincidence on the detectors in the lower arm of the setup means that 
the state of the modes $\hat{f}_x$ and $\hat{f}_y$ (before the beam
splitter BS2) has been collapsed
to the coherent superposition $(|0,2\rangle-|2,0\rangle)/\sqrt{2}$. This means that 
two photons must have been removed either from the mode $\hat{b}_x$ or
$\hat{b}_y$ by the beam splitter BS2. This gives the state of the remaining
photons of the form $|1,0;1,0\rangle + |0,1;0,1\rangle$, which has been defined
in Eq.~(\ref{Eq:Phi}).

We will now analyze the effect of imperfections of the auxiliary detectors
in the proposed scheme. The most critical issue is undercounting photons, when
the detectors give a fourfold single-photon coincidence even if more than
two photons have been reflected by the beam splitters BS1 and/or BS2. 
In such a case the output state will
be contaminated by additional terms containing a single photon or vacuum,
which are orthogonal to the maximally entangled two-photon component.
However, we shall demonstrate that the effect of detection losses can be suppressed
to some extent at the cost of decreasing the efficiency of the state preparation.

\begin{figure}[tbp]
  \centering
  \includegraphics[width=0.96\columnwidth]{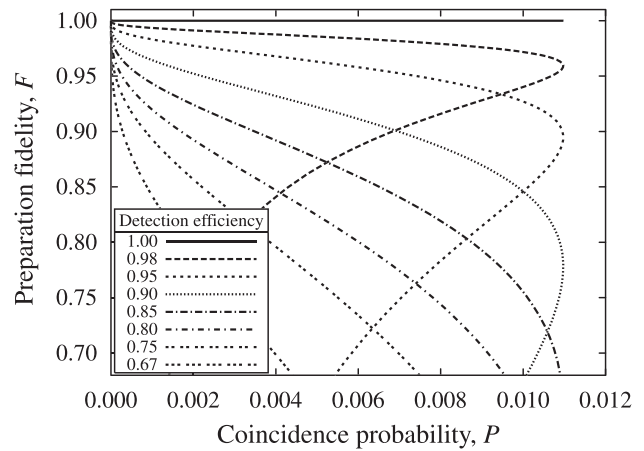}
  \caption{The fidelity $F$ of the prepared state versus the  probability
    $P$ of a fourfold single-photon coincidence for several values of
   the detection efficiency $\eta$. The curves are parametrized with
   the beam splitter coefficient $\theta$ running from $0$ to $\pi/2$.}
  \label{fig:fidplot}
\end{figure}

We shall model detector losses in the standard way assuming the same detection efficiency
equal to $\eta$ for all the four detectors. The most convenient way to include losses
in the previous calculations is to introduce additional tilded modes 
$\hat{\tilde{e}}_{x}, \hat{\tilde{e}}_{y}, \hat{\tilde{f}}_{x'}, \hat{\tilde{f}}_{y'}$
that are mixed
with the fields monitored by the detectors \cite{DetectionLosses}. This corresponds
to replacing the operator $\hat{e}_x$ in Eq.~(\ref{Eq:axassum}) by a superposition
$\sqrt{\eta}\hat{e}_x + \sqrt{1-\eta} \hat{\tilde{e}}_{x}$, and similarly for
$\hat{e}_y$, $\hat{f}_{x'}$, and $\hat{f}_{y'}$. The tilded operators 
$\hat{\tilde{e}}_{x},\ldots, \hat{\tilde{f}}_{y'}$ describe here
photons that escape detection due to nonunit efficiency. After straightforward
algebra we find that the
component of the reduced density matrix for the output modes
$(\hat{c}_x, \hat{c}_y; \hat{d}_x, \hat{d}_y)$ that is correlated with the
observation of a fourfold single-photon coincidence on the
trigger detectors has the following form:
\begin{eqnarray}
\label{Eq:ImperfectDetection}
\hat{\varrho} & = &
\frac{1}{4} \eta^4 \sin^8\theta[2\cos^4\theta |\Phi\rangle\langle\Phi|
+ (1-\eta)\sin^2\theta\cos^2\theta
\nonumber \\
& & \times (|1,0;0,0\rangle\langle 1,0;0,0|
+ |0,1;0,0\rangle\langle 0,1;0,0|
\nonumber \\
& & + |0,0;1,0\rangle\langle 0,0;1,0|
+ |0,0;0,1\rangle\langle 0,0;0,1|)
\nonumber \\
& & + 2 (1-\eta)^2 \sin^4\theta |\text{vac} \rangle \langle \text{vac} |],
\end{eqnarray}
where we have assumed for simplicity the transmission
coefficient to be equal for both the beam splitters: $\theta_a = \theta_b 
= \theta$, and the state $|\Phi\rangle$ has been defined in Eq.~(\ref{Eq:Phi}).
The different terms are added incoherently in the above formula,
as there is distinguishing information provided by the photons remaining
in the undetected tilded modes modeling losses.
We have kept the normalization of $\hat{\varrho}$
implied by the complete multimode wave function. This allows us to calculate
the probability of the coincidence event as the trace of $\hat{\varrho}$:
\begin{equation}
P = \text{Tr} \hat{\varrho} = \frac{1}{2} (\eta \sin^2 \theta)^4
(1-\eta \sin^2 \theta)^2,
\end{equation}
whereas the fidelity of the conditionally prepared state
reads
\begin{equation}
  F = \sqrt{\frac{ \langle \Phi | \hat{\varrho} |
\Phi \rangle}{\text{Tr} \hat{\varrho}}}
=
\frac{\cos^2\theta}{1-\eta\sin^2\theta}.
\end{equation}
In Fig.~\ref{fig:fidplot} we depict parametric plots $[P(\theta),F(\theta)]$
parametrized with the beam splitter coefficient $\theta$, for several
values of the detection efficiency $\eta$. All the curves span from the same point
$(P=0,F=1)$ corresponding to $\theta=0$, i.e., fully transmitting beam splitters,
to the maximum possible preparation efficiency [equal to the same value
$(2/9)^3$ as long as $\eta \ge 2/3$], with the fidelity of
the prepared state getting worse with decreasing $\eta$. However, it is clearly
seen in Fig.~\ref{fig:fidplot} that the
fidelity can be improved for any detection efficiency $\eta$ by increasing the beam splitter transmission. In this regime, the probability of reflecting more
than the minimum number of photons necessary to the trigger detectors becomes lower,
and consequently the danger of undercounting the auxiliary photons is less important.
This results in enhanced fidelity of the prepared state, though at the cost of
lower preparation efficiency. This effect demonstrates that the proposed scheme
is, in principle, robust to the effects of the
nonunit efficiency of the trigger detectors.

In conclusion, we have proposed
a conditional scheme based on a spontaneous parametric down-conversion
source that is capable of event-ready generation
of polarization entangled photon pairs. The scheme utilizes the three-pair
component of the down-conversion output, and the deleterious contribution
from the lower order term vanishes due to
the well-known destructive two-photon interference effect \cite{HongOuPRL87}.
The proposed scheme can be considered as a next-order
generalization of the earlier concept of entanglement swapping
\cite{ZukoZeilPRL93}.

As our scheme relies basically on polarization interference of photons
from one down-conversion source, the required experimental means
appear to be moderate. The critical parameters of the source are its
brightness and the preservation of the correlations in the number of produced
photons. These characteristics seem to be most promising for currently studied
sources based on nonlinear waveguides that offer improved control over the
spatio-temporal structure of the produced photons
\cite{PDCinFibers}. Assuming that the probability
of producing a single photon pair would be of the order of $5\%$, it is then 
easy to calculate that the probability of a fourfold coincidence in the proposed
setup is $7 \times 10^{-7}$. This should give a few dozens of useful photon pairs
per second for a pump laser with 100-MHz repetition rate, and this figure could
be improved by an order or two of magnitude by using a multi-gigahertz mode-locked
laser \cite{TomaOpL01},
provided that it can deliver enough pulse energy. These constraints, though
certainly challenging, do not seem to be very far from the capabilities of current
technology. It should also be noted that the visibility of the destructive
two-photon interference occurring for the second-order down-conversion term needs
to be high enough to make dominant the coincidence events generated by the next-order
term. The experimental observation of the destructive interference for the second-order
term should be feasible right now, as demonstrated by recent experiments by Lamas-Linares
{\em et al.} \cite{LamaHoweNAT01,HoweLamaPRL02}.

We acknowledge interesting discussions with I. A. Walmsley, A. K. Ekert,
and N. L\"{u}tkenhaus.
This research was supported by
ARO-administered MURI Grant No.\ DAAG-19-99-1-0125.
C.S.\ thanks European Science Foundation for a travel grant
under the Quantum Information Theory programme.

\end{document}